\newcommand{\bee}{\begin{equation}}
\newcommand{\ene}{\end{equation}}
\newcommand{\beea}{\begin{eqnarray}}
\newcommand{\enea}{\end{eqnarray}}

\documentclass[aps,prl,showpacs,twocolumn,superscriptaddress]{revtex4}
\usepackage{graphicx}
\usepackage{dcolumn}
\usepackage{bm}
\usepackage{amsmath}
\begin{document}
\title{Comment on "On “Novel attractive forces” between ions in quantum plasmas -- failure of linearized quantum hydrodynamics"}
\author{P. K. Shukla}
\affiliation{International Centre for Advanced Studies in Physical Sciences \& Institute for Theoretical Physics,
Faculty of Physics \& Astronomy, Ruhr University Bochum, D-44780 Bochum, Germany}
\affiliation{Department of Mechanical and Aerospace Engineering \& Center for Energy Research,
University of California San Diego, La Jolla, CA 92093}
\author{M. Akbari-Moghanjoughi}
\affiliation{Azarbaijan Shahid Madani University, Faculty of Sciences, Department of Physics, 51745-406 Tabriz, Iran}
\affiliation{International Centre for Advanced Studies in Physical Sciences, Ruhr University Bochum, D-44780 Bochum, Germany}
\author{B. Eliasson}
\affiliation{International Centre for Advanced Studies in Physical Sciences \& Institute for Theoretical Physics,
Faculty of Physics \& Astronomy, Ruhr University Bochum, D-44780 Bochum, Germany}
\begin{abstract}
In a recent paper Bonitz, Pehlke and Schoof \cite{Bonitz12}, hereafter referred to as BPS, have raised some points against the newly found
Shukla-Eliasson attractive potential \cite{Shukla12,Shukla12E}, hereafter refererred to as SEAP, around a stationary test charge in a quantum
plasma \cite{Shukla12,Shukla12E}. Our objective here is to discuss the inappropriateness of BPS reasoning concerning the applicability of the linearized quantum hydrodynamic theory, as well as to point out the shortcomings in BSP's arguments and to suggest for rescuing the BSP's
density functional theory and simulations which have failed to produce results that correctly match with that of Shukla and Eliasson \cite{Shukla12,Shukla12E}.
\end{abstract}
\pacs{52.30.-q,71.10.Ca, 05.30.-d}

\maketitle
Bonitz, Pehlke and Schoof \cite{Bonitz12} (BPS) have compared the findings from their density functional theory (DFT) simulations with that of an analytical theory of Shukla-Eliasson (SE) \cite{Shukla12,Shukla12E} which reported a short-range (at the scale size of several Bohr atomic radii) attractive potential (AP) distribution around a stationary ion test charge in a quantum plasma. To start with, we need to evaluate the assumptions in the quantum plasma regime employed in the theory of Shukla and Eliasson (SE). SE \cite{Shukla12} have used the well established \cite{Manfredi01, Manfredi05} quantum hydrodynamic (QHD) model for degenerate electrons to calculate the electric potential around a stationary ion test charge by using the electron dielectric constant deduced from the linearized electron continuity, nonrelativistic electron momentum and Poisson equations. The electron momentum equation includes the electrostatic and quantum forces (e.g the non-relativistic quantum statistical electron pressure in the zero-temperature free-electron Fermi gas limit, electron-exchange and electron correlation effects due to the electron spin, and the quantum electron recoil effect associated with overlapping of electron wave functions and their dispersion at atomic scales). Thus, the QHD model is subject to some approximations regarding quantum degeneracy of free electrons in a Fermi plasma and comparison of the quantum statistical pressure and the quantum recoil effect associated with electron tunneling through the quantum Bohm potential. The quantum electron degeneracy comes into the picture when the thermal de Broglie wavelength of electrons, $\lambda_B =\hbar/\sqrt{k_B T_p m_e}$  is comparable with the average inter-electron spacing $r_0 \approx (3/4\pi n_0)^{1/3}$, in addition to the condition $\lambda_B \ll \lambda_L$, where $\hbar$ is the Planck constant divided by $2\pi$, $k_B$ the Boltzmann constant, $T_p$ the plasma temperature, $m_e$ the rest mass of electrons, $n_0$ the electron number density, $\lambda_L =e^2/k_B T_p$ the Landau length \cite{Belyaev01}, and $e$ the magnitude of electron charge. However, the free electron assumption, which is implicitly hidden in the Fermi-Dirac statistical degeneracy pressure $P_{deg}$ \cite{Landau78}, is often overlooked in literature. It is important to note that the latter assumption is only valid for metallic compounds with very large electrical conductivity \cite{Kittel91}. This condition is also valid for solid density quantum plasmas and warm dense matter (WDM). Thus, the SEAP theory critically relies on the free-electron assumption, in addition to the zero-Fermi temperature electron degeneracy with $(\hbar\omega_p/k_B T_F)^2 \simeq 1$, and any application of this theory beyond those assumptions can lead to nonphysical results, as discussed below. Here $\omega_p =(4\pi n_0 e^2/m_e)^{1/2}$ is the electron plasma frequency and $T_F =(\hbar^2/2m_ek_B)(3\pi^2n_0)^{2/3}$ the Fermi electron temperature. The origin of the negative attractive SE potential is attributed to the consideration of the quantum recoil effect, which may dominate over the quantum statistical electron pressure as well as electron-exchange and electron-correlation effects, depending on the certain plasma density ranges discussed in Ref. \cite{Shukla12E}.

BPS have compared the SEAP theory with their DFT simulations for a hydrogen plasma. Their code reveals interesting mismatch between the two theories and therefore they concluded that the linearized QHD theory fails. Although the DFT theory is far from exact \cite{Perdew09} and still subject to many optimizations \cite{Chai04} with many different version of it found in the literature, one should seek the cause of fundamental divergences between the two approaches (the DFT and QHD models) in the inadequate comparison. BPS throughout their PRE paper \cite{Bonitz12} frequently refer to Friedel oscillations as a possible mechanism for the SE effect. We refute to such an assertion because, Friedel oscillations arise from localized perturbations in metallic and semiconducting materials and are absent in insulators with molecular bindings. Friedel oscillations are closely related to the Kohn anomaly \cite{Kohn59} caused by singularity in the plasma wave dispersion relation \cite{Lindhard54} (regardless of the quantum recoil effect at atomic scales) at the wavenumber value of $k=2k_F$, where $k_F$ is the Fermi wavenumber of electrons. Lindhard and Vidensk \cite{Lindhard54} have obtained the dielectric constant for the free electron liquid in metals, which becomes singular at some $k$ values. Although, such singularity is negligible in reciprocal space, however, its Fourier transform or its image in the real space causes a strong oscillations in the real-part of the dielectric function in proximity of the singularity, due to the well known Gibbs phenomenon. Such oscillations are reflected in the screening potential of the test charge in a Fermi electron liquid \cite{Simion05}. However, no such oscillations (Friedel oscillation of same origin) has been reported for quantum plasmas in which the Fermi-wavevector and the Fermi surface are not well-defined parameters. In quantum plasmas without the quantum recoil effect, one encounters only the Thomas-Fermi (TF) short-range [of the order of the TF screening radius $(k_B T_F/m_e)^{1/2}/\omega_p$] repulsive potential.

Now, let us closely inspect the points raised by BPS against the SEAP. In their Fig. 1, BPS \cite{Bonitz12} have shaded a region, they call the region of validity of the linearized QHD, for which $\hbar \omega_{p}< k_B T_{F}$. However, the QHD model is actually valid \cite{Crou08} for $(\hbar \omega_{p}/k_B T_F)^2 \sim 1$. Therefore, the shaded area in Fig. 1 of BPS does not apply to dense plasmas with zero-temperature completely degenerate electron fluids where electron-ion collisions are greatly inhibited by the Pauli blocking mechanism. This condition may be relaxed to
the quantum coupling parameter, $g_Q=(\hbar \omega_{p}/k_B T_F)^2$, even slightly greater than unity in the weak coupling limit \cite{Manfredi05}. In fact, the weak coupling limit in quantum plasmas is not a strict one and it closely relates to the mean-free-path of electrons which may be very large in the quantum plasma cases like some good conductors. In particular, it is well-known that in metallic compounds the characteristic plasmon frequency is at least one order of magnitude larger than that of electron-ion and many orders of magnitudes larger than that for electron-electron collisions \cite{Manfredi05}. As it will be apparent bellow, it is however irrelevant to apply the QHD degenerate free electron model used in SEAP theory \cite{Shukla12,Shukla12E} to nonmetallic hydrogen densities, characterized by the Brueckner parameter ($r_s=r_0/r_B$ with $r_B=\hbar^2/e^2m_e$ being the Bohr radius), lying in the region $g_Q >1$.

BPS in their Fig. 5 \cite{Bonitz12} compare their DFT simulations with SEAP for three different Brueckner parameters of $r_s=7,4,1.5$ corresponding to electron number densities of $n_0\simeq(4.7\times 10^{21},2.5\times 10^{22},4.8\times 10^{23})$ cm$^{-3}$, respectively. Since, the existence of the SEAP is a direct consequence of the interplay between the quantum forces which strongly rely on the free electron model and complete degeneracy assumptions, it is readily apparent that the SEAP does not apply for hydrogen with $r_s=7$ or even $r_s=4$. In fact, BPS correctly remark in p. 3
of their paper \cite{Bonitz12} that in the range $r_s>3$ hydrogen gas is in its nonmetallic molecular binding state and these bounds break only for $r_s\simeq 2\ldots 3$. That is, what they compare in their Fig. 5 for $r_s=4$ and $r_s=7$ is irrelevant for the SEAP, since it overrides the required assumptions for the applicability of the linearized QHD. In other words, the SEAP theory is an strongly ionized atomic theory applied only to highly conductive (pressure) ionized materials with densities beyond the Mott metal-insulator transition, which happens to be in the regime $r_s=1.2\ldots 1.5$ (e.g. see p. 3 of Ref. \cite{Bonitz12}). In fact, it is clearly observed from Fig. 5 of BPS that the best match between the two theories coincides with the value of $r_s=1.5$, which is close to the Mott transition for hydrogen composition. Therefore, it should be kept in mind that, despite the apparent wide density range for which the SEAP minimum exists in hydrogen plasma, the SEAP unlike the DFT is neither a theory for description of molecular bindings nor it gives rise to the molecular Lennard-Jones potential. Therefore, one must critically examine the applicability of the free electron assumption, which can only be valid for a metallic pressure-ionized hydrogen in the limit $r_s<1.2$. Obviously, this does not count as a defect for the SEAP theory, but, it is left for the reader to examine the outcome against the basic assumptions for the QHD model.

The Shukla-Eliasson attractive potential \cite{Shukla12}, which can lead ion-ion correlations, is caused by electron density localization due to quantum electron dispersion/recoil effect. It can be observed from Fig. 5 of BPS that for densities of $r_s=1\ldots 0.6$ the SEAP minimum can become more pronounced and the minimum of BPS disappears. Such defect in the standard DFT simulations may be overcome by extending the conventional Thomas-Fermi (TF) screening of ions to the modified Thomas-Fermi-Weizs\"{a}cker (TFW) improved model, which also includes the quantum electron recoil effect \cite{Chai04}. It has been shown (e.g. see Fig. 6 of Ref. \cite{Chai04}) that the improved TFW model can give rise to much deeper potential valley than that calculated with the ordinary TF model. The DFT theory with improved TFW density model has been also found to be more consistent with laboratory data compared to the standard DFT theories. Furthermore, BPS in their paper call the DFT simulation method as a reference to other theoretical results. This is by the way far from exact, since, there are major complexities associated with DFT approximate calculations \cite{Schuch09}, which should be overcome, before it can be claimed as a reference theory.

In summary, the material presented in BPS paper is partially (where related to molecular binding) irrelevant to the SEAP theory. On the other hand, BPS in their PRE paper \cite{Bonitz12} have erroneously attributed the existence of the SEAP minimum to the Friedel oscillations (related to the singularity in the Lindhard dielectric constant for ordered Fermi liquids possessing sharp Fermi surface), relevant to metals and semiconductors with a well defined Fermi-wavevector. The SEAP minimum has a direct root in the electron quantum recoil effect and electron wave-function interferences, present in any quantum plasma in the absence of a well-defined Brillouin zone. Moreover, the free-electron model assumption used in the SEAP theory has been overlooked by BPS in their PRE paper. The DFT theory, in its current state, is far from being complete in the context of plasma physics with an ensemble of degenerate electrons interacting in a collective fashion at atomic dimensions, and many improvements are under way and until then one is unable to claim DFT as an ultimate theory. Finally, the extent of validity of the linear hydrodynamic description of ion structure factors in dense plasmas has been investigated by Mithen {\it et al.} \cite{mithen} who concluded that such an approach can be used to effectively model the ion response in compressed plasmas for a wide range of the plasma number densities that can be probed experimentally.

\end{document}